# Thermal stability of monolayer $WS_2$ in BEOL conditions

Simona Pace[1, 2, *], Marzia Ferrera[3], Domenica Convertino[1, 2], Giulia Piccinini[1, 4], Michele Magnozzi[3, 5], Neeraj Mishra[1, 2], Stiven Forti[1], Francesco Bisio[6], Maurizio Canepa[3, 5], Filippo Fabbri[1, #], Camilla Coletti[1, 2, *]

[1] Center for Nanotechnology Innovation CNI@NEST, Istituto Italiano di Tecnologia, Piazza San Silvestro 12, 56127 Pisa

[2] Graphene Labs, Istituto Italiano di Tecnologia, Via Morego 30, 16163 Genova

[3] OPTMATLAB, Dipartimento di Fisica, Università degli Studi di Genova, via Dodecaneso 33, 16146 Genova, Italy

[4] NEST, Scuola Normale Superiore, Piazza San Silvestro 12, 56127 Pisa, Italy

[5] Istituto Nazionale di Fisica Nucleare, Via Dodecaneso 33, 16146 Genova, Italy

[6] CNR-SPIN, Corso Perrone 24, 16152 Genova, Italy

[#] Present address: NEST, Istituto Nanoscienze – CNR, Scuola Normale Superiore, Piazza San Silvestro 12, 56127 Pisa, Italy

[*] Corresponding authors: simona.pace@iit.it, camilla.coletti@iit.it

## Abstract

Monolayer tungsten disulfide ($WS_2$) has recently attracted large interest as a promising material for advanced electronic and optoelectronic devices such as photodetectors, modulators, and sensors. Since these devices can be integrated in a silicon (Si) chip via back-end-of-line (BEOL) processes, the stability of monolayer $WS_2$ in BEOL fabrication conditions should be studied. In this work, the thermal stability of monolayer single-crystal $WS_2$ at typical BEOL conditions is investigated; namely (i) heating temperature of 300 °C, (ii) pressures in the medium- ($10^{-3}$ mbar) and high- ($10^{-8}$ mbar) vacuum range; (iii) heating times from 30 minutes to 20 hours. Structural, optical and chemical analyses of $WS_2$ are performed via scanning electron microscopy (SEM), Raman spectroscopy, photoluminescence (PL) and X-ray photoelectron spectroscopy (XPS). It is found that monolayer single-crystal $WS_2$ is intrinsically stable at these temperature and pressures, even after 20 hours of thermal treatment. The thermal stability of $WS_2$ is also preserved after exposure to low-current electron beam (12 pA) or low-fluence laser (0.9 mJ/$\mu m^2$), while higher laser fluencies cause photo-activated degradation upon thermal treatment. These results are instrumental to define fabrication and in-line monitoring procedures that allow the integration of $WS_2$ in device fabrication flows without compromising the material quality.

# Introduction

Monolayer transition metal dichalcogenides (TMDs) have attracted great interest due to their excellent electrical and optical properties [1–8]: because of their semiconducting nature, TMDs allow to overcome the shortcomings of zero-bandgap graphene, showing attractive potential for constructing digital circuits [9,10] and next generation light-emitting devices [11,12]. In addition to the large range of available bandgaps of pristine S-, Se- and Te-based TMDs [13–17], the facile intermixing in different alloys [18–22] and stacking in heterostructures [23–25] give way to virtually infinite possibilities of electronic and optical properties engineering. Moreover, the integration of large-area semiconducting TMDs with existing complementary metal oxide silicon (CMOS) platforms promises fast advances in electronic and photonic technologies [26–29]. In CMOS fabrication procedures, silicon (Si) chips are processed with a series of steps grouped into two fabrication portions known as front-end-of-line (FEOL) and back-end-of-line (BEOL) [30]. In the FEOL individual devices are patterned on the Si chips, while in the BEOL these devices are interconnected with each other and with the outside world. Similarly to graphene, TMDs can be monolithically integrated with well-established CMOS BEOL [28]. Indeed, TMDs-Si chips have already been reported for a number of applications [29,31–35]. This compatibility gives to TMDs a great advantage in terms of process optimization and cost reductions with respect to other "beyond-CMOS" candidates, such as III-V semiconductors or Ge, in which indirect integration with Si chips is much more challenging [36].

Monolayer tungsten disulfide ($WS_2$) has attracted large interest due to its unique properties, such as large spin–orbit splitting at the valence band K-point (462 meV) [37] and high emission quantum yield [38]. By manipulating the spin and valley degrees of freedom of $WS_2$, novel spintronic and valleytronic devices can be developed [39,40]. $WS_2$ also shows remarkably large light-matter interaction with high exciton binding energy (700 meV) [41] that, together with the possibility to grow large-area monolayer, have made $WS_2$ an enticing candidate for applications in electronics [29], optoelectronics [42] and photonics [34,43]. Indeed, some examples of $WS_2$-based devices fabricated in typical CMOS BEOL conditions have already been reported: Yang and co-workers [34] have demonstrated a $WS_2$-based all-optical modulator fully integrated with typical CMOS $Si_3N_4$ waveguides to modulate a 532 nm pump light source. Moreover, a fully CMOS-compatible graphene-hBN-$WS_2$ metal-insulator-semiconductor transistor showing ambipolar behaviour has also been recently reported [35]. These early results confirm that the optimization of $WS_2$-Si chips can be boosted by employing already-optimized CMOS BEOL fabrication steps, rather than implementing a completely new protocol.

In light of the great applicative potential of monolayer $WS_2$, its stability under typical BEOL processing conditions needs to be assessed to devise correct handling and fabrication procedures. To date, it has been reported that – under ambient conditions – monolayer $WS_2$ slowly reacts with environmental oxygen and fully degrades within 1 year [44]. Instead, if the environmental aging takes place at higher temperatures, the degradation accelerates [45,46]. Rong et al. [46] found that polycrystalline WS$_2$ monolayer starts to degrade in static air after 90 minutes at 250 °C and after only 20 minutes at 380 °C [46]. Additionally, it has been observed that the substrate on which WS$_2$ is grown can play a role in either boosting [47] or hindering [48] the oxidation of WS$_2$, depending on the electron transfer between WS$_2$ and the substrate [47]. Experimental and theoretical works agree on the fact that WS$_2$ degradation initiates at defective sites (i.e. sulfur vacancies, grain boundaries, edges) and it is due to the oxidation of WS$_2$ when interacting with environmental oxide species.

The abovementioned studies shed light on the stability and the degradation mechanism of WS$_2$ in air, at either room temperature or upon annealing. However, typical BEOL processes, such as deposition of dielectrics, evaporation of metal contacts and post-deposition annealings, are carried out at temperatures between 300 °C and 400 °C and base pressures that range from a few mTorr (~$10^{-3}$ mbar) to $10^{-6}$ mTorr (~$10^{-9}$ mbar) [49–53]. Therefore, the results already present in literature do not explore the stability of monolayer WS$_2$ under BEOL conditions. To fill this gap, in this work monolayer WS$_2$ is exposed to standardized conditions that are representative of typical BEOL fabrication protocols [28,54,55], and its structural, chemical and optical properties are studied. In addition to the fabrication steps, BEOL procedures also include inline monitoring of the material quality by adopting fast, non-contact methods, such as optical and scanning electron (SEM) microscopies and Raman and photoluminescence (PL) spectroscopies [27]. A fast, non-destructive microscopic technique is indeed necessary to evaluate the morphological properties of the fabricated sample with suitable resolution and without any required sample preparation that would destroy the device. In this sense, both optical and SEM imaging can be good candidates, depending on the size of the single-crystals. Recently, it has been suggested that the degradation of WS$_2$ monolayer at room temperature can be photo-induced by exposure to focused light and that the rate of this phenomenon is strictly dependent on the power of the light used [56,57]. Hence, before integrating WS$_2$ in BEOL flows, the compatibility of WS$_2$ to intermediate characterization steps involving techniques based on focused lights, such as Raman and PL spectroscopies, should be first demonstrated and the optimal conditions, needed to preserve the quality of the material, found.

In this work, monolayer WS$_2$ was epitaxially grown on hydrogen-etched sapphire substrate ($\alpha$-Al$_2$O$_3$ (0001)) via low pressure chemical vapor deposition (LP-CVD) and its thermal stability was evaluated at typical BEOL conditions (300 °C and base pressure of $10^{-3}$ mbar and $10^{-8}$ mbar). The morphological, structural and optical modifications upon annealing were evaluated and it was found that freshly grown WS$_2$ monolayer is intrinsically stable in these conditions even for long annealing times (up to 20 hours). Furthermore, the stability of WS$_2$ was investigated at 300 °C and medium-vacuum (1 x $10^{-3}$ mbar) after exposure to commonly used characterization techniques, i.e. SEM, Raman, X-ray photoemission (XPS) and PL spectroscopies. It was observed that SEM (e-beam current 12 pA) has no effect on the stability of WS$_2$ at 300 °C, while the exposure to laser can trigger the degradation of WS$_2$. In particular, if WS$_2$ is exposed to a 532 nm laser with a fluence of 9.4 mJ/µm$^2$ (laser power 730 µW), the morphological and optical properties of WS$_2$ rapidly degrade upon annealing at 300 °C. If the laser fluence is below a threshold of 0.9 mJ/μm$^2$ (laser power 73 μW), instead, no oxidation is initiated and monolayer WS$_2$ remains stable at 300 °C for several hours. These results are instrumental not only for future successful integration of monolayer WS$_2$ in BEOL fabrications, but also to avoid sample damaging when performing fundamental analyses that rely on laser adoption and sample annealing.

## Experimental Methods

### Sample preparation

The substrates used for CVD growth of monolayer WS$_2$ were dice cut from c-axis, HEMCOR single-crystal, sapphire α-Al$_2$O$_3$ (0001) wafers supplied by Alfa Aesar (Germany). Before WS$_2$ growth, sapphire substrates were cleaned via sonication in acetone, isopropanol, and de-ionized (DI) water, then immersed in piranha solution (1:3, H$_2$O$_2$:H$_2$SO$_4$) for 15 min and finally washed in DI water. Subsequently, the dice were etched in hydrogen atmosphere as described in reference [58] to remove polishing scratches and reveal atomic steps.

The etched sapphire substrate was then directly loaded in the CVD reactor for the growth of monolayer WS$_2$ via low-pressure CVD. Tungsten trioxide (WO$_3$, powder, Sigma Aldrich, 99.995%) and sulfur (S, pellets, Sigma Aldrich, 99.998%) were used as solid precursors. The process was performed within a 2.5-inches horizontal hot-wall furnace (Lenton PTF): the furnace comprises a central hot-zone (growth-zone), where a crucible loaded with WO$_3$ powder was placed 3 cm away from the growth substrate, and an inlet zone, where the S powder was positioned and separately heated by a resistive belt at 120 °C. Temperature in the growth zone was concurrently ramped-up to 930 °C, at a chamber pressure of ~5 x 10$^{-2}$ mbar. During the growth argon was adopted as carrier gas, as described in references [37,59].

### WS$_2$ thermal annealing

Medium-vacuum thermal annealing of WS$_2$ was performed in a hot-wall CVD quartz tube reactor, while high-vacuum annealing was carried out in an ultra-high-vacuum (UHV) chamber used for XPS measurements. In the CVD reactor, the annealings were performed at a temperature of 300 °C, a base pressure of 1 x 10$^{-3}$ mbar and a temperature ramp-up rate of 5 °C/min. The medium-vacuum annealing was carried out either in a continuative manner up to 10 hours or by extracting and characterizing the sample after each annealing step. The high-vacuum annealing was performed at 300 °C, at a base pressure of 2 x 10$^{-8}$ mbar for 10 hours, consecutively.

### Characterization techniques

Raman and PL analyses were performed using a Renishaw InVia system equipped with a 532 nm green laser, a 100× objective lens (0.89 NA) and a spot size of ~1 μm. All Raman and PL experiments were carried out in standard laboratory conditions (temperature of 22 °C, 30% humidity) and at atmospheric pressure. Before and after each analysis, the samples were kept in dark and in low vacuum (1 mbar). Different laser powers were used to evaluate the influence of laser exposure on the thermal stability of monolayer $WS_2$. All single spectra were acquired at the center of the same crystal and, unless otherwise stated, the laser power was set at 730 μW or 73 μW and the exposure time was 10 s. The fluence was calculated to be 9.4 mJ/μm$^2$ and 0.9 mJ/μm$^2$, respectively. SEM analysis was carried out using an In-Lens detector, with an accelerating voltage of 2 keV and an electron current of 12 pA (specimen current < 0.5 pA) in order to minimize the charging of the insulating substrate and the damaging due to the electron beam irradiation. Low energy electron diffraction (LEED) measurements were performed in ultra-high-vacuum (UHV) with a SPECS Er-LEED optics. The electron beam energy was kept above 150 eV to avoid charging from the sample. XPS was performed within a PHI-instruments 5800 station equipped with a monochromatized Al K$_α$ X-ray source. The X-ray spot size on the sample was few-hundred micrometers. The energy resolution was set to 100 meV for high-resolution spectra. The binding energy scale was calibrated by setting the adventitious carbon C1s peak at 284.8 eV. The W4f and S2p peaks were fitted by means of an approximated Voigt function (Gaussian/Lorentzian product form, mixing = 0.5, chosen from CasaXPS line-shapes database).

## Results and Discussion

### 1. Growth and characterization of monolayer CVD $WS_2$

Monolayer $WS_2$ was grown on hydrogen-etched α-$Al_2O_3$ (0001) [58] using low-pressure CVD following the procedure described in the Experimental section. Figure 1a shows a typical secondary electron (SE) micrograph of the synthesized $WS_2$ single-crystals with a lateral average size of 5 μm. The darker SE contrast, obtained with the In-Lens detector, is due to the attenuation of the secondary electrons emerging from the substrate beneath the layer of $WS_2$ [60]. The symmetry of the $Al_2O_3$ (0001) surface imposes a registry on which $WS_2$ crystals are observed to grow with rotation of 30 ° ± 60 ° with respect to the crystal lattice of the underlying α-$Al_2O_3$. This mutual orientation is clearly visible in the SEM image in Figure 1a and the percentage of misoriented crystals is estimated to be 30% over a dataset of 161 single-crystals. A low energy electron diffraction (LEED) pattern measured over an area of about 1 mm$^2$ is shown in Figure 1b and further confirms the epitaxial alignment of the synthesized $WS_2$ crystals with respect to the substrate. The pattern perfectly matches the superimposed model (blue and red dots), which represents $WS_2$ R30 (blue circles) over the $Al_2O_3$ (0001) surface (red circles). No pattern of the Al-rich ($\sqrt{31}\times\sqrt{31}$)$R \pm 9$ surface reconstruction is instead visible after $WS_2$ growth [58]. The displayed LEED pattern also exhibits some diffused intensity, which cannot be ascribed neither to the substrate, nor to the $WS_2$. The diffused intensity is in fact most likely due to unreacted amorphous $WO_3$ or to some transient chemical compound, as also suggested by XPS results (see Figure 4). Minor rotational disorder, due to the presence of few misoriented $WS_2$ single-crystals over the large area analyzed, is also visualized in the LEED pattern as a very faint continuous ring of intensity, crossing the main $WS_2$ reflections, as suggested also by SEM results (Figure 1a).

Raman mapping (Figure 1c) and spectroscopy (Figure 1d) confirm that WS$_2$ is monolayer with a high degree of homogeneity. The benchmarking for WS$_2$ monolayer is that the ratio of the 2LA(M) + E$_{2g}$($\Gamma$) and A$_{1g}$($\Gamma$) Raman modes is higher than 2.2, as previously demonstrated in reference [61]. It is worth noting that the increase of the 2LA(M) + E$_{2g}$($\Gamma$) /A$_{1g}$($\Gamma$) ratio on the edge of the crystal, visible in Figure 1c, is an intrinsic artifact of high resolution Raman mapping. When the laser reaches the edge of the crystal, there is a concurrent decrease of all mode intensities and the intensity of the A$_{1g}$($\Gamma$) goes close to noise level, causing an increase of the 2LA(M) + E$_{2g}$($\Gamma$) /A$_{1g}$($\Gamma$) ratio. PL measurements (Figure 1e and f) also confirm that the synthesized WS$_2$ is monolayer: the room-temperature PL spectrum, reported in Figure 1e, shows an intense emission at (630.7 ± 0.1) nm (1.96 eV), related to the exciton of monolayer WS$_2$. The PL peak is slightly red-shifted if compared to the PL peak position typically observed for WS$_2$ on sapphire (~ 620 nm) [57,62–64]. Moreover, this peak has a strong asymmetry, showing a tail on the high wavelength side. These results are likely due to the presence of intra-gap states related to sulfur vacancies, as reported by Carozo et al. [65]. The PL map (Figure 1f) shows an increase of intensity at the edges of the crystal, while the center shows homogeneous PL signal, as it was also previously reported [66,67]. The statistic distributions of 2LA(M) + E$_{2g}$($\Gamma$) /A1g($\Gamma$) ratio, A1g($\Gamma$) Raman shift, PL intensity and PL position are reported in Figure S1.

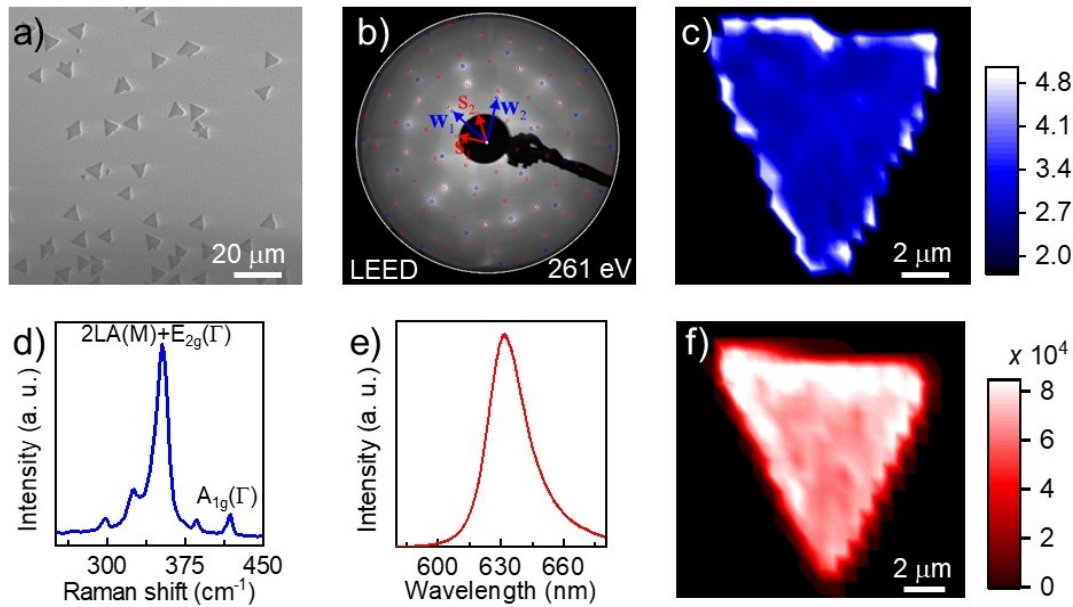

**Figure 1 Characterization of WS$_2$ grown on $\alpha$-Al$_2$O$_3$.** a) SEM image of oriented WS$_2$ single-crystals grown on hydrogen-etched $\alpha$-Al$_2$O$_3$. b) LEED pattern of WS$_2$ on $\alpha$-Al$_2$O$_3$ showing the relative orientation of WS$_2$ R30 (W$_1$ and W$_2$ in blue) and Al$_2$O$_3$ (0001) c-plane (S$_1$ and S$_2$ in red). c) Map of the ratio of the 2LA + E$_{2g}$ / A$_{1g}$ Raman modes. d) Representative Raman spectrum of WS$_2$, the 2LA + E$_{2g}$ and A$_{1g}$ peaks used in panel c are highlighted. e) Representative PL spectrum of WS$_2$. f) PL map of WS$_2$ showing the intense emission form the A-exciton at (630.7 ± 0.1) nm (1.96 eV).

## 2. Thermal stability of monolayer WS$_2$

To investigate the stability of WS$_2$ in typical BEOL conditions, freshly grown WS$_2$ single-crystals were initially annealed at a temperature of 300 °C and medium-vacuum (pressure of 10$^{-3}$ mbar). These conditions were chosen to be as close as possible to those at which 2D materials are usually exposed during device fabrication steps [50,51], i.e. dielectric deposition, metal evaporation or post-deposition annealing. To further mimic typical device fabrication flows, in which the sample is subjected to multiple heatings during the different processing steps, the annealing (cumulatively 10 hours long) was interrupted several times (i.e., after 30 minutes, 1 hour, 3 hours and 10 hours) and the sample characterized after each step by means of SEM, Raman and PL spectroscopies, which have been suggested as inline metrology for TMD-Si BEOL flow [27] (details in the Experimental section). This intermediate characterization, indeed, is often carried out to monitor the quality of the material and of the assembling device during its fabrication. In Figure 2 the characterization of WS$_2$ after sequential annealing steps is reported. From SEM micrographs (Figure 2a – c) a change of shape of the crystal is visible already after 3 hours of annealing (i.e., 30' + 30' + 2h), suggesting that the irregular edges of the as-grown WS$_2$ (Figure 2a) are more reactive than the center of the single-crystal, in agreement with previous results [47,68]. After 10 hours annealing (i.e., 30' + 30' + 2h + 7h) (Figure 2c) small triangular holes are also visible both on the edges and in inner areas of the crystal (white triangles), indicating loss of WS$_2$ material. Additional SEM micrographs taken on the same sample after the fifth and sixth annealing steps, for a total of 15 and 20 hours, are reported in Figure S2a and b. Triangular holes, already visible on the edge of the crystal after 10 hours annealing, become increasingly larger (i.e., lateral size up to 500 nm) and denser on the entire crystal, with a decrease of the overall area of WS$_2$ monolayer from ∼ 81 μm$^2$ (as grown) down to ∼ 74 μm$^2$ (20 hours annealing).

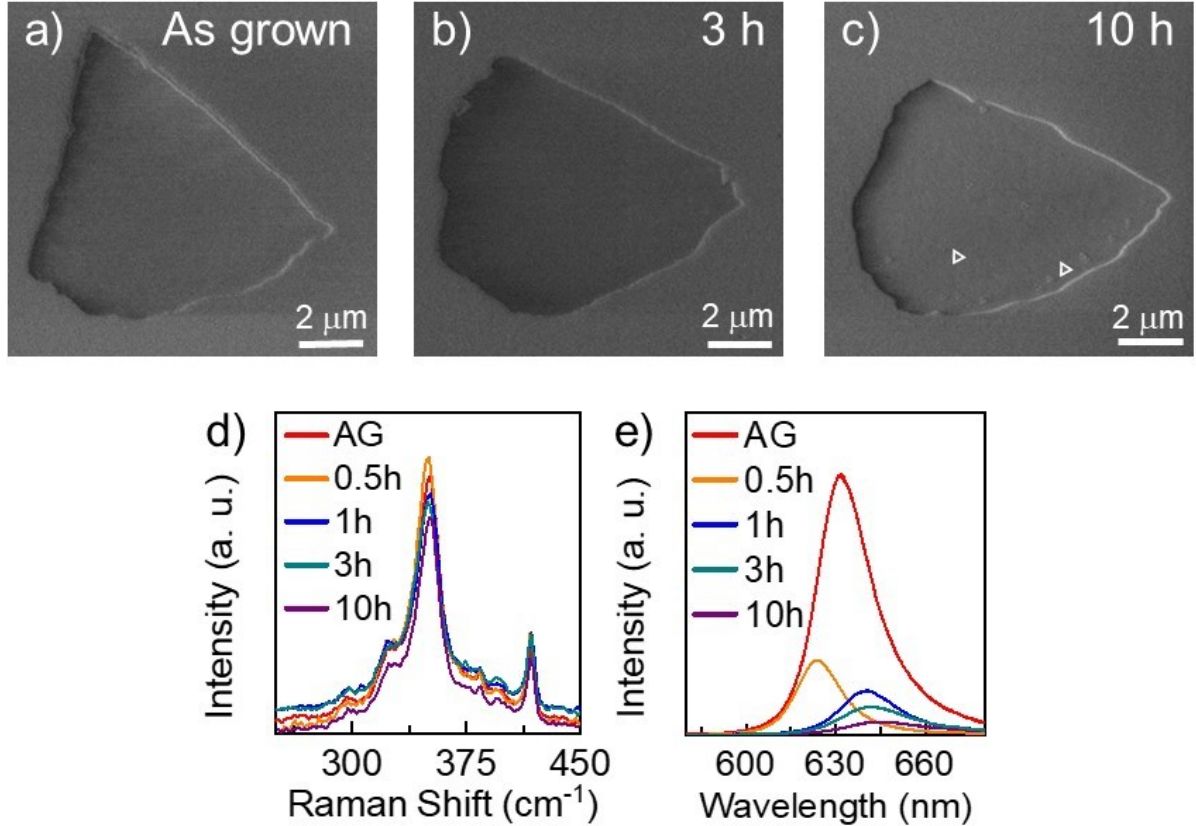

**Figure 2 Effect of multiple annealings of $WS_2$ at 300 °C and $10^{-3}$ mbar.** a – c) SEM micrographs of $WS_2$ as grown and after 3 steps (cumulative 3h) and 4 steps (cumulative 10h) annealing at 300 °C. d – e) Raman (d) and PL (e) spectra of $WS_2$ after multiple annealing at 300 °C taken after each step.

Figure 2d shows the evolution of the Raman spectrum after multiple annealing steps from 30 minutes up to 10 hours, carried out through 4 sequential steps. The annealing in these conditions mainly affects the intensity of the 2LA(M) + $E_{2g}(\Gamma)$ Raman mode: as shown in Figure 2d, after a first slight increase (30 minutes), the 2LA(M) + $E_{2g}(\Gamma)$ intensity slowly decreases for the subsequent steps. Raman spectra taken after additional annealing steps of 15 and 20 cumulative hours are reported in Figure S2c, in which a large decrease of the 2LA(M) + $E_{2g}(\Gamma)$ peak intensity is visible, down to a minimum of 52% of the starting value (i.e., after 20 hours). The decrease of the intensity of the 2LA(M) + $E_{2g}(\Gamma)$ mode for annealings longer than 1 hour is related to the formation of holes in the $WS_2$ crystal (Figure a - c) and therefore to a loss of the overall quantity of material analyzed under the laser spot. The PL spectra for each annealing step are reported in Figure 2e. The PL peak related to the $WS_2$ A exciton before annealing lies at (630.7 ± 0.1) nm (1.96 eV) with a full width at half maximum (FWHM) of (14 ± 1) nm. Upon annealing, the PL intensity rapidly decreases and becomes negligible after 15 hours at 300 °C (Figure S2d). After the first annealing step, the PL blue shifts down to (623 ± 0.2) nm (1.99 eV), while with increasing annealing time, it red-shifts up to (644.2 ± 0.2) nm (1.93 eV) with a concurrent broadening (FWHM (29 ± 1) nm). The first blue-shift of PL can be attributed to residual contaminations deposited on the growth substrate desorbing during the first annealing step. The subsequent broadening and red shift of the PL peak is likely due to a higher population of point defects such as sulfur vacancies (e.g., at the edge of the triangular holes), a hypothesis that is supported by several works [65,68,69]. The PL peak broadening is also combined with an increase of peak asymmetry. This asymmetry is due to the presence of an additional luminescence peak at about 650 nm, which is related to sulfur vacancy bound excitons [65,67].

From these preliminary results, one might conclude that monolayer $WS_2$ is not stable at 300 °C in medium-vacuum and that its properties rapidly degrade after the first few annealing steps. However, this large instability at relatively low temperature and pressure is in contrast with both the melting and the dissociation temperatures of bulk $WS_2$, which are around 1250 °C [70] and 1040 °C [71], respectively. Also, such degradation is more dramatic than that reported by Rong and co-workers [46], who observed that after 20 minutes of annealing at 380 °C *in air*, only highly defective areas of polycrystalline $WS_2$ start to degrade. The fast degradation of monolayer single-crystal $WS_2$ reported in Figure 2 might then have been accelerated by external events and can have different origins: (i) the generation of defects upon exposure to laser and/or electron beam during the intermediate characterization analyses; (ii) the presence of a relatively high concentration of residual oxide species in the chamber during the annealing at the vacuum level employed; (iii) the exposure to air after each annealing step.

To shed light on this, each of the possible degradation causes was systematically studied. First, the role of intermediate characterization after each annealing step was evaluated by analyzing two different areas, namely 1 and 2, of a fresh $WS_2$ sample using SEM, Raman and PL spectroscopies, respectively. In Figure 3a (area 1) two $WS_2$ crystals (black arrows) were imaged using only SEM with e-beam current 12 pA (specimen current < 0.5 pA); otherwise the sample was left in dark. The $WS_2$ crystal reported in Figure 3c (area 2, white arrow) was additionally characterized using Raman and PL spectroscopies with laser power set to 730 μW (fluence 9.4 mJ/μm$^2$), as in Figure 2. The values of both e-beam current for SEM imaging and laser fluence for Raman and PL spectroscopies were chosen to be compatible with typical values used to characterize TMDs on insulating substrate, i.e. sapphire. Higher e-beam current would lead to charging of the sample during SEM analysis, while higher laser fluence can damage the exposed material. The sample was then annealed at 300 °C and 10$^{-3}$ mbar for 10 hours in a continuative manner. While no difference is visible in the morphology of the crystal in area 1 before and after annealing (Figure 3a and b), a large morphological degradation is observed for the $WS_2$ crystal in area 2, which was exposed to a laser power of 730 μW (white arrow, Figure 3c and d). Moreover, both Raman and PL signals of the analyzed crystal in area 2 largely decrease after the continuative annealing at 300 °C (Figure 3e and f). These results suggest that the values of e-beam current typically used for SEM on sapphire preserve the intrinsic stability of $WS_2$ at 300 °C and medium-vacuum (10$^{-3}$ mbar), even if heated for 10 continuative hours. Conversely, the previous exposure of $WS_2$ single-crystal to a focused laser (~1 μm) with fluence 9.4 mJ/μm$^2$ appears to trigger the instability of WS$_2$, which then degrades upon annealing.

To further investigate the relation between laser fluence and thermal instability of WS$_2$, an additional area (area 3) was characterized using Raman and PL spectroscopies with laser power set to 73 μW (fluence 0.9 mJ/μm$^2$) prior the annealing at 300 °C and 10$^{-3}$ mbar for 10 hours (Figure 3g – l). Panels 3g – j report the SEM micrographs and the Raman and PL spectra of WS$_2$ exposed to laser power of 73 μW (fluence 0.9 mJ/μm$^2$), before and after annealing: no clear difference in the morphology nor in Raman or PL signal is visible. These results indicate that WS$_2$ is intrinsically stable at 300 °C in medium-vacuum as long as the degradation is not triggered by the exposure to laser with a fluence higher than a certain threshold (9.4 mJ/μm$^2$). On the contrary, below this threshold, the thermal stability is preserved.

It is interesting to note that, even when the degradation of WS$_2$ is triggered by exposing the single-crystal to a relatively high-power laser, this phenomenon is localized only to the exposed area, as shown in Figure 3d. After the annealing, the previously exposed crystal (white arrow) is nearly destroyed, while a small WS$_2$ crystal next to it (Figure 3d, up left corner) appears to be fully preserved. These results also rule out that oxidative species in the chamber might be responsible for the degradation observed in Figure 2 (hypothesis (ii)), as both Raman and PL of WS$_2$ (Figure 3i and j) are unaffected after a 10-hours annealing in the chamber. It is also worth mentioning that the different levels of degradation for WS$_2$, exposed to a laser power of 730 μW (fluence 9.4 mJ/μm$^2$) and annealed for 10 hours in a multiple (Figure 2c) and continuative (Figure 3d) manner, are expected to be related only to the different mapping methods adopted in the two experiments (see Figure S3 for further details).

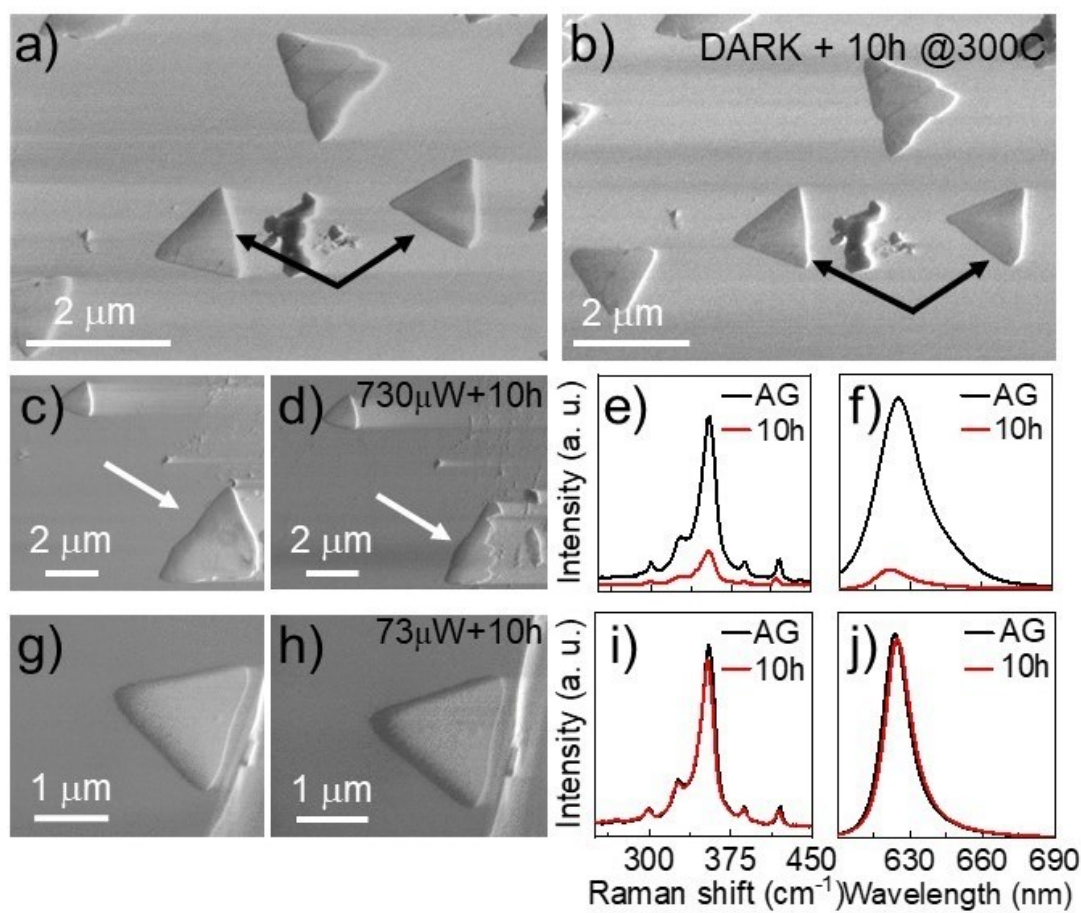

**Figure 3 Analysis of the influence of intermediate characterization on the stability of WS$_2$ at 300 °C.** a – b) SEM micrographs of WS$_2$ exposed only to SEM beam before (a) and after (b) annealing at 300 °C. c – d) SEM micrographs of WS$_2$ exposed to both SEM beam and green laser with laser power of 730 μW before (c) and after (d) annealing at 300 °C. e - f) Raman (e) and PL (f) spectra before and after annealing of sample shown in panel c and d. g – h) SEM micrographs of WS$_2$ exposed to both SEM beam and green laser with laser power of 73 μW before (g) and after (h) annealing at 300 °C. i - j) Raman (i) and PL (j) spectra before and after annealing of sample shown in panel g and h.

In Figure 4, the role of the pressure in the chamber is taken under consideration by annealing WS$_2$ at $10^{-8}$ mbar (high-vacuum) for 10 hours: again, no large difference after the annealing is observed if the sample is only analyzed by means of SEM with e-beam current set to 12 pA (panel a and b), as typically used for insulating substrates, and laser fluence of 0.9 mJ/μm$^2$ (panels c and d). Both Raman and PL spectra, reported in panel c and d respectively, confirm that using low laser fluence is safe for WS$_2$: both structural and optical properties of WS$_2$ are seemingly preserved after the annealing in high-vacuum. Indeed, although the PL intensity in panel d slightly decreases after annealing, the FWHM of the WS$_2$ A-exciton peak before and after the thermal treatment remains ~12 nm with no redshift, indicating that the optical properties of WS$_2$ are not considerably degraded by the treatment. As the annealing was performed in a chamber adopted for XPS analysis, it was also possible to perform chemical characterization of the as-grown sample and monitor the effect that such a long annealing (i.e., 10 hours) has on WS$_2$ chemical composition. The plots in Figure 4e show the high-resolution spectra, after background subtraction, acquired within the binding energy (BE) window of tungsten W4f and sulfur S2p core electrons, before and after the continuative high-vacuum annealing, respectively. In order to study the chemistry of the sample surface, a deconvolution of the experimental curves was performed. The tungsten spectrum of the as-grown sample can be fitted as the superposition of three doublets, with the W4f$_{7/2}$ components at (32.1 ± 0.1) eV (yellow), (32.9 ± 0.1) eV (light blue) and (36.0 ± 0.1) eV (pink), respectively. The W5p$_{3/2}$ peaks are shifted 5.8 eV above the corresponding W4f$_{7/2}$. Both the yellow and light-blue W4f doublets at lower BE have their counterpart in the sulfur spectrum, which shows two doublets with S2p$_{3/2}$ components at (161.7 ± 0.1) eV and (162.5 ± 0.1) eV. The yellow and light-blue peaks are ascribed to atoms belonging to WS$_2$ crystals, while the pink doublet was ascribed to residuals of unreacted WO$_3$ precursor. In particular, the energy position of the light-blue subcomponent is in perfect agreement with the one reported in literature for the hexagonal WS$_2$ semiconducting phase (2H), while the lower BE of about 0.8 eV related to the yellow minor subcomponent can be a fingerprint of the 1T-WS$_2$ metallic phase, sulfur vacancies or other kinds of defects due to CVD growth process [45,72,73]. Interestingly, the spectra acquired after 10 hours continuative annealing at 300 °C in high-vacuum show almost unaltered yellow and light-blue components. The only significant change in the W spectrum is related to the WO$_3$ doublet (due to unreacted material on the sapphire surface) [37], which shows a decreased contribution to the W4f envelope in comparison to the pre-annealing data. Moreover, a new W-related doublet (green) appears at (33.6 ± 0.1) eV, which does not have a counterpart in the S2p spectrum acquired after the annealing. The green component can be ascribed to a chemical evolution into lower valence oxides of the unreacted WO$_3$ [74] as a consequence of the high-vacuum annealing, which, however, does not have any sizeable effect on the thermal stability of WS$_2$ crystals.

The absence of modifications in the surface chemistry of WS$_2$ crystals, confirmed by XPS results, is also particularly important since desorption of sulfur in a BEOL could cause line contamination.

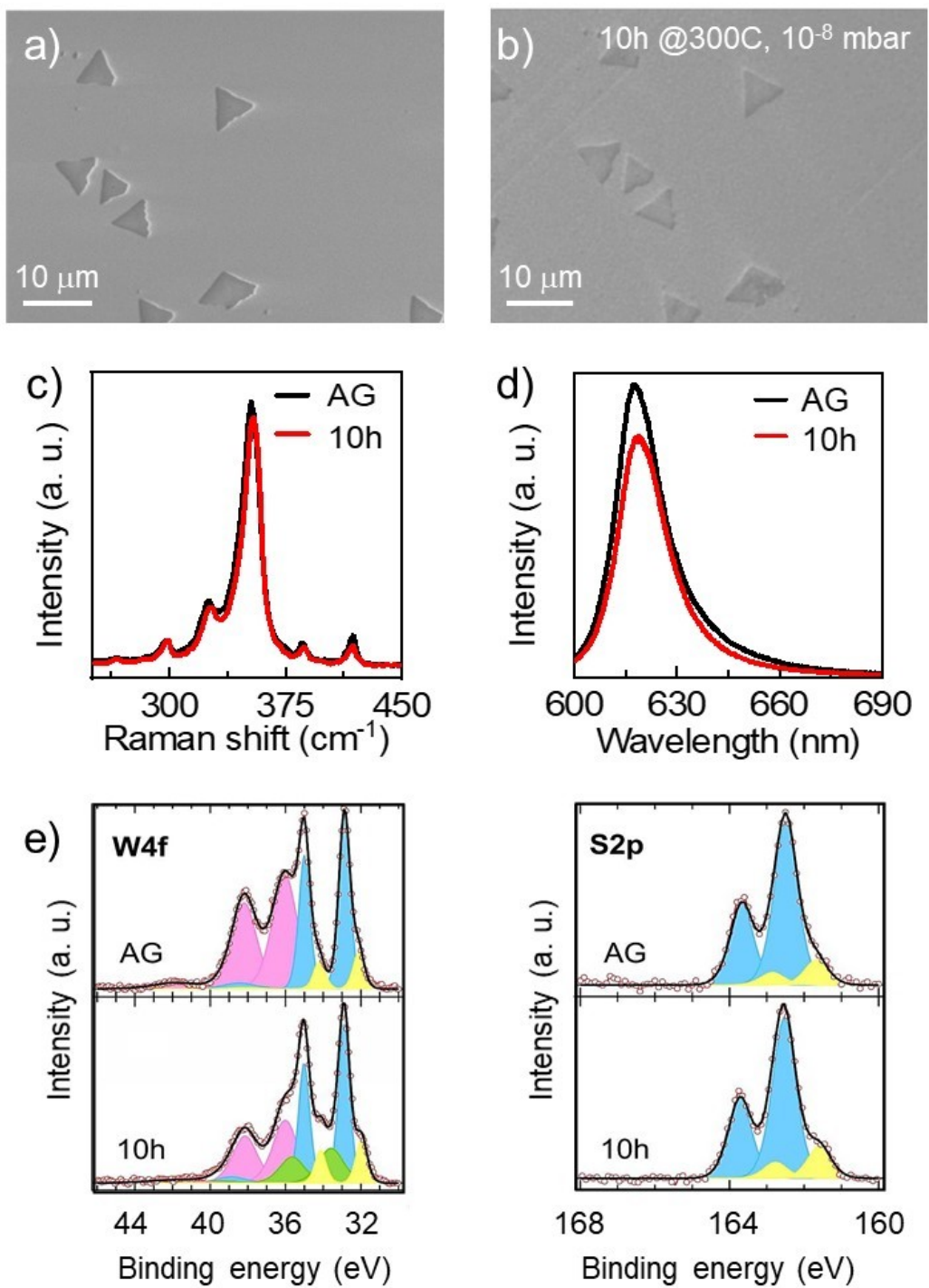

**Figure 4 Analysis of the thermal stability of WS$_2$ at 300 °C in high-vacuum (10$^{-8}$ mbar).** a – b) SEM micrographs of WS$_2$ before (a) and after (b) annealing at 300 °C. c – d) Raman (c) and PL (d) spectra before and after annealing of the sample shown in panel a and b. e) XPS high-resolution W4f and S2p experimental spectra (markers) and fitting results (black line) before and after 10 hours continuative high-vacuum annealing.

To investigate whether the exposure of the sample to air after each annealing step (listed as cause (iii)) might contribute to the fast degradation reported in Figure 2, a fresh sample was annealed at 300 °C in medium-vacuum in 2 subsequent steps of 10 hours each and analyzed using the safe conditions found in Figure 3. No clear difference is observed from the SEM, Raman and PL analyses on the same crystal (Figure S4), confirming that, if no oxidation is triggered by a relatively high-power laser, $WS_2$ is intrinsically stable at 300 °C and medium-vacuum even when exposing the sample to air in-between multiple annealings.

Finally, the thermal stability of $WS_2$ at high temperatures was also investigated by annealing a fresh sample of $WS_2$ at 600 °C and $10^{-3}$ mbar. Figure S5 shows that at this annealing temperature $WS_2$ becomes unstable and complete degrades after 1 hour. The XPS spectra of a sample after 2 hours annealing under the above-mentioned conditions show a complete absence of $WS_2$ related components and that traces of tungsten oxides are the only remaining species (Figure S5e).

### 3. Compatibility of monolayer $WS_2$ with typical BEOL conditions

The results reported in this work indicate that monolayer single-crystal $WS_2$ is intrinsically stable at 300 °C, a typical temperature adopted in BEOL fabrications steps, even for 10 hours. However, care must be taken in order to preserve its thermal stability and the appropriate conditions for handling the sample between each step must be carefully chosen. A schematic summary of the conditions investigated in this work is reported in Figure 5. If a base pressure of $10^{-3}$ mbar (or lower) is used when annealing at 300 °C, $WS_2$ does not degrade: the vacuum level is low enough to avoid fast oxidation of $WS_2$ at a relatively high temperature.

The inline monitoring of the sample in-between annealing steps needs to be carefully performed. In fact, monolayer $WS_2$ remains stable at 300 °C and $10^{-3}$ mbar as long as the degradation is not triggered by exposure to a focused light with relatively high power, in agreement with previous reports [56,57]. The exposure to laser above a certain threshold either generates or activates defects, i.e. sulfur vacancies, that then rapidly react at high temperature with oxidative species that are still present in the annealing chamber at a pressure of $10^{-3}$ mbar. However, if the laser fluence is lowered down to $\leq 0.9$ mJ/$\mu m^2$ (laser power 73 µW), this photoinduced process is not activated. Atkin and collaborators [56] reported a fluence threshold of defect photo-activation for $WS_2$ between 2 and 20 mJ/$\mu m^2$ at room temperature [56]. They suggest that if the laser exposure is energetic enough the top sulfur layer of $WS_2$ is affected, generating localized defects (i.e. contaminant adsorption), and becomes more susceptible to deterioration; below the threshold, instead, the reaction is not activated. Our experiments indicate that carrying out analysis with a light fluence of 0.9 mJ/$\mu m^2$ is safe for monolayer $WS_2$, even when annealing at 300 °C and in medium-vacuum. Moreover, multiple exposures to the laser with light fluence of 0.9 mJ/$\mu m^2$ and different combinations of laser power and exposure time (i.e. laser power 730 µW, exposure time 1 s, fluence 0.9 mJ/$\mu m^2$) were taken into consideration and no major difference was observed.

On the contrary, SEM analysis with an e-beam current of 12 pA does not induce degradation of $WS_2$ single-crystals. A relation between e-beam current and defect activation of $WS_2$ might be expected, similarly to the laser fluence. In perspective of BEOL integration on a Si platform, particular care should be adopted and the effect of higher electron currents on monolayer $WS_2$ should be investigated.

Finally, the effect of ambient exposure in-between processing steps was investigated by annealing $WS_2$ at 300 °C in two subsequent steps of 10 hours each. It was found that $WS_2$ is compatible with sequential fabrication steps at this temperature and in medium-vacuum.

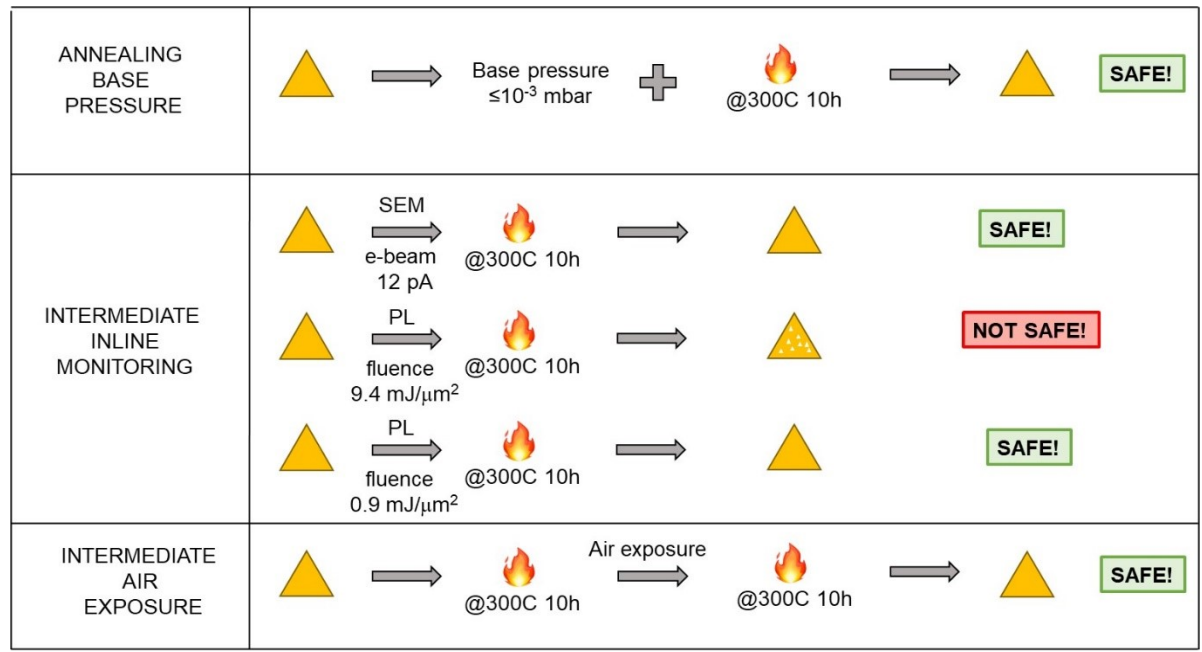

**Figure 5 Schematic summary of the thermal stability of $WS_2$ in the conditions studied in this work.** At a chamber pressure $\leq 10^{-3}$ mbar, the vacuum level is high enough to prevent oxidation of $WS_2$ at 300 °C. Intermediate inline monitoring of $WS_2$ using SEM with an e-beam current of 12 pA or PL using fluence $\leq 0.9$ mJ/μm² are compatible with annealing $WS_2$ at 300 °C and medium-vacuum. PL with a laser fluence equal to 9.4 mJ/μm² instead induces defects. Intermediate exposure to air of $WS_2$ in-between annealing steps is compatible with the thermal stability of $WS_2$ at 300 °C and medium-vacuum.

## Conclusions

In summary, in this work the thermal stability of monolayer single-crystal $WS_2$ at 300 °C in medium- ($10^{-3}$ mbar) and high- ($10^{-8}$ mbar) vacuum was investigated. These conditions were chosen to be as representative as possible to typical BEOL conditions, at which $WS_2$ would be exposed if integrated in wafer-scale device fabrication flows. Epitaxial monolayer $WS_2$ grown on sapphire was first annealed in these conditions via 4 sequential steps for a total time of 10 hours. After each step, the sample was unloaded and fully analyzed by means of SEM, Raman and PL spectroscopies, in a way that can mimic the fabrication & control method typically used in device fabrication procedures. It was found that, if high laser fluence (power 730 μW, fluence 9.4 mJ/μm²) was used for the spectroscopic analyses, $WS_2$ rapidly degraded in these thermal conditions with a large deterioration of its chemical and structural properties, resulting in a large quenching of its optical properties. On the contrary, if a laser fluence of 0.9 mJ/μm$^2$ was used, monolayer WS$_2$ did not degrade upon annealing at 300 °C and medium- or high-vacuum. From these results, it can be concluded that WS$_2$ can undergo processing steps involving a temperature of 300 °C (in a non-reactive atmosphere) in medium- and high-vacuum, provided that the intermediate characterization steps are carried out compatibly with the photo-activation threshold of WS$_2$ oxidation. This compatibility can be achieved by either working with a laser fluence below the photo-activation threshold or by choosing a sacrificial crystal on which the characterization is carried out.

## Acknowledgements

The research leading to these results has received funding from the European Union's Horizon 2020 research and innovation program under grant agreement no. 785219-GrapheneCore2 and n. 881603-GrapheneCore3 and from Compagnia di San Paolo (project STRATOS).

## References

[1] Khanna V K, *2016*, Transition Metal DichalcogenidesTransition Metal Dichalcogenides -Based Nanoelectronics, pp 313–22

[2] Ovchinnikov D, Allain A, Huang Y S, Dumcenco D and Kis A, *2014*, Electrical transport properties of single-layer WS$_2$, *ACS Nano* **8** 8174–81

[3] Magnozzi M, Ferrera M, Piccinini G, Pace S, Forti S, Fabbri F, Coletti C, Bisio F and Canepa M, *2020*, Optical dielectric function of two-dimensional WS$_2$ on epitaxial graphene, *2D Mater.* **7** 025024

[4] Manzeli S, Ovchinnikov D, Pasquier D, Yazyev O V. and Kis A, *2017*, 2D transition metal dichalcogenides, *Nat. Rev. Mater.* **2** 17033

[5] Chhowalla M, Liu Z and Zhang H, *2015*, Two-dimensional transition metal dichalcogenide (TMD) nanosheets, *Chem. Soc. Rev.* **44** 2584–6

[6] Zhou H, Wang C, Shaw J C, Cheng R, Chen Y, Huang X, Liu Y, Weiss N O, Lin Z, Huang Y and Duan X, *2015*, Large area growth and electrical properties of p-type WSe$_2$ atomic layers, *Nano Lett.* **15** 709–13

[7] Obeid M M, Stampfl C, Bafekry A, Guan Z, Jappor H R, Nguyen C V., Naseri M, Hoat D M, Hieu N N, Krauklis A E, Vu T V. and Gogova D, *2020*, First-principles investigation of nonmetal doped single-layer BiOBr as a potential photocatalyst with a low recombination rate, *Phys. Chem. Chem. Phys.* **22** 15354–64

[8] Abed Al-Abbas S S, Muhsin M K and Jappor H R, *2019*, Two-dimensional GaTe monolayer as a potential gas sensor for SO$_2$ and NO$_2$ with discriminate optical properties, *Superlattices Microstruct.* **135** 106245

[9] Wang H, Yu L, Lee Y H, Shi Y, Hsu A, Chin M L, Li L J, Dubey M, Kong J and Palacios T, *2012*, Integrated circuits based on bilayer MoS$_2$ transistors, *Nano Lett.* **12** 4674–80

[10] Radisavljevic B, Whitwick M B and Kis A, *2011*, Integrated circuits and logic operations based on single-layer MoS$_2$, *ACS Nano* **5** 9934–8

[11] Salehzadeh O, Tran N H, Liu X, Shih I and Mi Z, *2014*, Exciton kinetics, quantum efficiency, and

efficiency droop of monolayer MoS$_2$ light-emitting devices, *Nano Lett.* **14** 4125–30

[12]   Jo S, Ubrig N, Berger H, Kuzmenko A B and Morpurgo A F, *2014*, Mono- and bilayer WS$_2$ light-emitting transistors, *Nano Lett.* **14** 2019–25

[13]   Gutiérrez H R, Perea-López N, Elías A L, Berkdemir A, Wang B, Lv R, López-Urías F, Crespi V H, Terrones H and Terrones M, *2013*, Extraordinary room-temperature photoluminescence in triangular WS$_2$ monolayers, *Nano Lett.* **13** 3447–54

[14]   Conley H J, Wang B, Ziegler J I, Haglund R F, Pantelides S T and Bolotin K I, *2013*, Bandgap engineering of strained monolayer and bilayer MoS$_2$, *Nano Lett.* **13** 3626–30

[15]   Tongay S, Zhou J, Ataca C, Lo K, Matthews T S, Li J, Grossman J C and Wu J, *2012*, Thermally driven crossover from indirect toward direct bandgap in 2D Semiconductors: MoSe$_2$ versus MoS$_2$, *Nano Lett.* **12** 5576–80

[16]   Mak K F, Lee C, Hone J, Shan J and Heinz T F, *2010*, Atomically thin MoS$_2$: A new direct-gap semiconductor, *Phys. Rev. Lett.* **105** 136805

[17]   Ruppert C, Aslan O B and Heinz T F, *2014*, Optical properties and band gap of single- and few-layer MoTe$_2$ crystals, *Nano Lett.* **14** 6231–6

[18]   Abdulraheem Z and Jappor H R, *2020*, Tailoring the electronic and optical properties of SnSe$_2$/InS van der Waals heterostructures by the biaxial strains, *Phys. Lett. Sect. A Gen. At. Solid State Phys.* **384** 126909

[19]   Obeid M M, Shukur M M, Edrees S J, Khenata R, Ghebouli M A, Khandy S A, Bouhemadou A, Jappor H R and Wang X, *2019*, Electronic band structure, thermodynamics and optical characteristics of BeO$_{1-x}$A$_x$ (A = S, Se, Te) alloys: Insights from ab initio study, *Chem. Phys.* **526** 110414

[20]   Xie L M, *2015*, Two-dimensional transition metal dichalcogenide alloys: Preparation, characterization and applications, *Nanoscale* **7** 18392–401

[21]   Zhang W, Li X, Jiang T, Song J, Lin Y, Zhu L and Xu X, *2015*, CVD synthesis of Mo$_{(1-x)}$W$_x$S$_2$ and MoS$_{2(1-x)}$Se$_{2x}$ alloy monolayers aimed at tuning the bandgap of molybdenum disulfide, *Nanoscale* **7** 13554–60

[22]   Meng Y, Wang T, Li Z, Qin Y, Lian Z, Chen Y, Lucking M C, Beach K, Taniguchi T, Watanabe K, Tongay S, Song F, Terrones H and Shi S F, *2019*, Excitonic Complexes and Emerging Interlayer Electron-Phonon Coupling in BN Encapsulated Monolayer Semiconductor Alloy: WS$_{0.6}$Se$_{1.4}$, *Nano Lett.* **19** 299–307

[23]   Okada M, Kutana A, Kureishi Y, Kobayashi Y, Saito Y, Saito T, Watanabe K, Taniguchi T, Gupta S, Miyata Y, Yakobson B I, Shinohara H and Kitaura R, *2018*, Direct and Indirect Interlayer Excitons in a van der Waals Heterostructure of hBN/WS$_2$/MoS$_2$/hBN, *ACS Nano* **12** 2498–505

[24]   Ross J S, Rivera P, Schaibley J, Lee-Wong E, Yu H, Taniguchi T, Watanabe K, Yan J, Mandrus D, Cobden D, Yao W and Xu X, *2017*, Interlayer Exciton Optoelectronics in a 2D Heterostructure p-n Junction, *Nano Lett.* **17** 638–43

[25]   Sahoo P K, Memaran S, Nugera F A, Xin Y, Díaz Márquez T, Lu Z, Zheng W, Zhigadlo N D, Smirnov D, Balicas L and Gutiérrez H R, *2019*, Bilayer Lateral Heterostructures of Transition-Metal Dichalcogenides and Their Optoelectronic Response, *ACS Nano* **13** 12372–84

[26]   Sun Z, Martinez A and Wang F, *2016*, Optical modulators with 2D layered materials, *Nat. Photonics* **10** 227–38

[27] Akinwande D, Huyghebaert C, Wang C-H, Serna M I, Goossens S, Li L-J, Wong H-S P and Koppens F H L, *2019*, Graphene and two-dimensional materials for silicon technology, *Nature* **573** 507–18

[28] Neumaier D, Pindl S and Lemme M C, *2019*, Integrating graphene into semiconductor fabrication lines, *Nat. Mater.* **18** 525–9

[29] Schram T, Smets Q, Groven B, Heyne M H, Kunnen E, Thiam A, Devriendt K, Delabie A, Lin D, Lux M, Chiappe D, Asselberghs I, Brus S, Huyghebaert C, Sayan S, Juncker A, Caymax M and Radu I P, *2017*, WS$_2$ transistors on 300 mm wafers with BEOL compatibility, *European Solid-State Device Research Conference* (IEEE) pp 212–5

[30] Rinerson D and Cheung R, *2012*, Device Fabrication Method

[31] Rodder M A, Vasishta S and Dodabalapur A, *2020*, Double-Gate MoS$_2$ Field-Effect Transistor with a Multilayer Graphene Floating Gate: A Versatile Device for Logic, Memory, and Synaptic Applications, *ACS Appl. Mater. Interfaces* **12** 33926–33

[32] Yim C, McEvoy N, Riazimehr S, Schneider D S, Gity F, Monaghan S, Hurley P K, Lemme M C and Duesberg G S, *2018*, Wide Spectral Photoresponse of Layered Platinum Diselenide-Based Photodiodes, *Nano Lett.* **18** 1794–800

[33] Ansari L, Monaghan S, McEvoy N, Coileáin C, Cullen C P, Lin J, Siris R, Stimpel-Lindner T, Burke K F, Mirabelli G, Duffy R, Caruso E, Nagle R E, Duesberg G S, Hurley P K and Gity F, *2019*, Quantum confinement-induced semimetal-to-semiconductor evolution in large-area ultra-thin PtSe$_2$ films grown at 400 °C, *npj 2D Mater. Appl.* **3** 33

[34] Yang S, Liu D C, Tan Z L, Liu K, Zhu Z H and Qin S Q, *2018*, CMOS-Compatible WS$_2$-Based All-Optical Modulator, *ACS Photonics* **5** 342–6

[35] Lee G, Oh S, Kim J and Kim J, *2020*, Ambipolar Charge Transport in Two-Dimensional WS$_2$ Metal-Insulator-Semiconductor and Metal-Insulator-Semiconductor Field-Effect Transistors, *ACS Appl. Mater. Interfaces* **12** 23127–33

[36] Kazior T E, *2014*, Beyond CMOS: heterogeneous integration of III–V devices, RF MEMS and other dissimilar materials/devices with Si CMOS to create intelligent microsystems, *Philos. Trans. R. Soc. A Math. Phys. Eng. Sci.* **372** 20130105

[37] Forti S, Rossi A, Büch H, Cavallucci T, Bisio F, Sala A, Menteş T O, Locatelli A, Magnozzi M, Canepa M, Müller K, Link S, Starke U, Tozzini V and Coletti C, *2017*, Electronic properties of single-layer tungsten disulfide on epitaxial graphene on silicon carbide, *Nanoscale* **9** 16412–9

[38] Yuan L and Huang L, *2015*, Exciton dynamics and annihilation in WS$_2$ 2D semiconductors, *Nanoscale* **7** 7402–8

[39] Benítez L A, Savero Torres W, Sierra J F, Timmermans M, Garcia J H, Roche S, Costache M V. and Valenzuela S O, *2020*, Tunable room-temperature spin galvanic and spin Hall effects in van der Waals heterostructures, *Nat. Mater.* **19** 170–5

[40] Garcia J H, Cummings A W and Roche S, *2017*, Spin hall effect and weak antilocalization in graphene/transition metal dichalcogenide heterostructures, *Nano Lett.* **17** 5078–83

[41] Zhu B, Chen X and Cui X, *2015*, Exciton Binding Energy of Monolayer WS$_2$, *Sci. Rep.* **5** 9218

[42] Lan C, Zhou Z, Zhou Z, Li C, Shu L, Shen L, Li D, Dong R, Yip S and Ho J C, *2018*, Wafer-scale synthesis of monolayer WS$_2$ for high-performance flexible photodetectors by enhanced chemical vapor

deposition, *Nano Res.* **11** 3371–84

[43] Datta I, Chae S H, Bhatt G R, Li B, Yu Y, Cao L, Hone J and Lipson M, *2018*, Giant electro-refractive modulation of monolayer $WS_2$ embedded in photonic structures, *Conference on Lasers and Electro-Optics* (Washington, D.C.: OSA) p STu4N.7

[44] Gao J, Li B, Tan J, Chow P, Lu T-M and Koratkar N, *2016*, Aging of Transition Metal Dichalcogenide Monolayers, *ACS Nano* **10** 2628–35

[45] Perrozzi F, Emamjomeh S M, Paolucci V, Taglieri G, Ottaviano L and Cantalini C, *2017*, Thermal stability of $WS_2$ flakes and gas sensing properties of $WS_2/WO_3$ composite to $H_2$, $NH_3$ and $NO_2$, *Sensors Actuators B Chem.* **243** 812–22

[46] Rong Y, He K, Pacios M, Robertson A W, Bhaskaran H and Warner J H, *2015*, Controlled Preferential Oxidation of Grain Boundaries in Monolayer Tungsten Disulfide for Direct Optical Imaging, *ACS Nano* **9** 3695–703

[47] Fabbri F, Dinelli F, Forti S, Sementa L, Pace S, Piccinini G, Fortunelli A, Coletti C and Pingue P, *2020*, Edge Defects Promoted Oxidation of Monolayer $WS_2$ Synthesized on Epitaxial Graphene, *J. Phys. Chem. C* **124** 9035–44

[48] Kang K, Godin K, Kim Y D, Fu S, Cha W, Hone J and Yang E-H, *2017*, Graphene-Assisted Antioxidation of Tungsten Disulfide Monolayers: Substrate and Electric-Field Effect, *Adv. Mater.* **29** 1603898

[49] English C D, Shine G, Dorgan V E, Saraswat K C and Pop E, *2016*, Improved contacts to $MoS_2$ transistors by ultra-high vacuum metal deposition, *Nano Lett.* **16** 3824–30

[50] Late D J, Liu B, Matte H S S R, Dravid V P and Rao C N R, *2012*, Hysteresis in single-layer $MoS_2$ field effect transistors, *ACS Nano* **6** 5635–41

[51] Vervuurt R H J, Kessels W M M E and Bol A A, *2017*, Atomic Layer Deposition for Graphene Device Integration, *Adv. Mater. Interfaces* **4** 1700232

[52] Andric S, Ohlsson Fhager L, Lindelöw F, Kilpi O-P and Wernersson L-E, *2019*, Low-temperature back-end-of-line technology compatible with III-V nanowire MOSFETs, *J. Vac. Sci. Technol. B* **37** 061204

[53] Bishop M D, Hills G, Srimani T, Lau C, Murphy D, Fuller S, Humes J, Ratkovich A, Nelson M and Shulaker M M, *2020*, Fabrication of carbon nanotube field-effect transistors in commercial silicon manufacturing facilities, *Nat. Electron.* **3** 492–501

[54] Kozhakhmetov A, Torsi R, Chen C Y and Robinson J A, *2020*, Scalable Low-Temperature Synthesis of Two-Dimensional Materials Beyond Graphene, *J. Phys. Mater.* **4** 012001

[55] Huyghebaert C, Schram T, Smets Q, Kumar Agarwal T, Verreck D, Brems S, Phommahaxay A, Chiappe D, El Kazzi S, Lockhart de la Rosa C, Arutchelvan G, Cott D, Ludwig J, Gaur A, Sutar S, Leonhardt A, Marinov D, Lin D, Caymax M, Asselberghs I, Pourtois G and Radu I P, *2018*, 2D materials: roadmap to CMOS integration, *2018 IEEE International Electron Devices Meeting (IEDM)* vol 2018-Decem (IEEE) pp 22.1.1-22.1.4

[56] Atkin P, Lau D W M, Zhang Q, Zheng C, Berean K J, Field M R, Ou J Z, Cole I S, Daeneke T and Kalantar-Zadeh K, *2018*, Laser exposure induced alteration of $WS_2$ monolayers in the presence of ambient moisture, *2D Mater.* **5** 015013

[57] Kotsakidis J C, Zhang Q, Vazquez De Parga A L, Currie M, Helmerson K, Gaskill D K and Fuhrer M S, *2019*, Oxidation of Monolayer $WS_2$ in Ambient Is a Photoinduced Process, *Nano Lett.* **19** 5205–15

[58] Mishra N, Forti S, Fabbri F, Martini L, McAleese C, Conran B R, Whelan P R, Shivayogimath A, Jessen B S, Buß L, Falta J, Aliaj I, Roddaro S, Flege J I, Bøggild P, Teo K B K and Coletti C, *2019*, Wafer-Scale Synthesis of Graphene on Sapphire: Toward Fab-Compatible Graphene, *Small* **15** 1904906

[59] Rossi A, Spirito D, Bianco F, Forti S, Fabbri F, Büch H, Tredicucci A, Krahne R and Coletti C, *2018*, Patterned tungsten disulfide/graphene heterostructures for efficient multifunctional optoelectronic devices, *Nanoscale* **10** 4332–8

[60] Kochat V, Nath Pal A, Sneha E S, Sampathkumar A, Gairola A, Shivashankar S A, Raghavan S and Ghosh A, *2011*, High contrast imaging and thickness determination of graphene with in-column secondary electron microscopy, *J. Appl. Phys.* **110** 014315

[61] Berkdemir A, Gutiérrez H R, Botello-Méndez A R, Perea-López N, Elías A L, Chia C-I, Wang B, Crespi V H, López-Urías F, Charlier J-C, Terrones H and Terrones M, *2013*, Identification of individual and few layers of $WS_2$ using Raman Spectroscopy, *Sci. Rep.* **3** 1755

[62] Zhang Y, Zhang Y, Ji Q, Ju J, Yuan H, Shi J, Gao T, Ma D, Liu M, Chen Y, Song X, Hwang H Y, Cui Y and Liu Z, *2013*, Controlled growth of high-quality monolayer $WS_2$ layers on sapphire and imaging its grain boundary, *ACS Nano* **7** 8963–71

[63] Cong C, Shang J, Wang Y and Yu T, *2018*, Optical Properties of 2D Semiconductor $WS_2$, *Adv. Opt. Mater.* **6** 1700767

[64] Xu Z Q, Zhang Y, Lin S, Zheng C, Zhong Y L, Xia X, Li Z, Sophia P J, Fuhrer M S, Cheng Y B and Bao Q, *2015*, Synthesis and Transfer of Large-Area Monolayer $WS_2$ Crystals: Moving Toward the Recyclable Use of Sapphire Substrates, *ACS Nano* **9** 6178–87

[65] Carozo V, Wang Y, Fujisawa K, Carvalho B R, McCreary A, Feng S, Lin Z, Zhou C, Perea-López N, Elías A L, Kabius B, Crespi V H and Terrones M, *2017*, Optical identification of sulfur vacancies: Bound excitons at the edges of monolayer tungsten disulfide, *Sci. Adv.* **3** e1602813

[66] Gutiérrez H R, Perea-López N, Elías A L, Berkdemir A, Wang B, Lv R, López-Urías F, Crespi V H, Terrones H and Terrones M, *2013*, Extraordinary room-temperature photoluminescence in triangular $WS_2$ monolayers, *Nano Lett.* **13** 3447–54

[67] Okuno Y, Lancry O, Tempez A, Cairone C, Bosi M, Fabbri F and Chaigneau M, *2018*, Probing the nanoscale light emission properties of a CVD-grown $MoS_2$ monolayer by tip-enhanced photoluminescence, *Nanoscale* **10** 14055–9

[68] Fabbri F, Rotunno E, Cinquanta E, Campi D, Bonnini E, Kaplan D, Lazzarini L, Bernasconi M, Ferrari C, Longo M, Nicotra G, Molle A, Swaminathan V and Salviati G, *2016*, Novel near-infrared emission from crystal defects in $MoS_2$ multilayer flakes, *Nat. Commun.* **7** 13044

[69] McCreary A, Berkdemir A, Wang J, Nguyen M A, Elías A L, Perea-López N, Fujisawa K, Kabius B, Carozo V, Cullen D A, Mallouk T E, Zhu J and Terrones M, *2016*, Distinct photoluminescence and Raman spectroscopy signatures for identifying highly crystalline $WS_2$ monolayers produced by different growth methods, *J. Mater. Res.* **31** 931–44

[70] Anon, *1994*, *Concise Encyclopedia Chemistry* (Berlin, New York: DE GRUYTER)

[71] Brainard W A, *1969*, Thermal stability and friction of disulfides, diselenides, and ditellurides of molybdenum and tungsten in ultrahigh vacuum, *NASA; United States, Tech. reports*

[72] Donarelli M, Prezioso S, Perrozzi F, Bisti F, Nardone M, Giancaterini L, Cantalini C and Ottaviano L,

*2015*, Response to $NO_2$ and other gases of resistive chemically exfoliated $MoS_2$-based gas sensors, *Sensors Actuators, B Chem.* **207** 602–13

[73]   Donarelli M, Bisti F, Perrozzi F and Ottaviano L, *2013*, Tunable sulfur desorption in exfoliated $MoS_2$ by means of thermal annealing in ultra-high vacuum, *Chem. Phys. Lett.* **588** 198–202

[74]   Xie F Y, Gong L, Liu X, Tao Y T, Zhang W H, Chen S H, Meng H and Chen J, *2012*, XPS studies on surface reduction of tungsten oxide nanowire film by $Ar^+$ bombardment, *J. Electron Spectros. Relat. Phenomena* **185** 112–8